\documentclass[preprint,aps,showpacs]{revtex4}
\usepackage[dvips]{graphics}
\usepackage{amssymb}
\usepackage{epsfig}
\usepackage{amsmath}
\usepackage{makeidx}

\makeindex

\begin{document}

\title{Polarization dynamics in quantum dots: The
role of dark excitons.}
\author{E. Tsitsishvili}
\address{Institute for Cybernetics, Tbilisi Technical University,  S. Euli 5,
0186,  Tbilisi Georgian Republic}
\author{H. Kalt}
\address{Institut f\"ur Angewandte Physik, Karlsruhe Institute of Technology (KIT), D-76128
Karlsruhe, Germany} \draft

\begin{abstract}

We study an impact of the fine structure of the heavy--hole ground
state exciton confined in semiconductor quantum dots on the
photoluminescence polarization dynamics solving the relevant
system of the rate equations.  The presence of the dark excitons
is usually ignored and the polarization decay is assumed to be
caused by direct transitions within the radiative doublet. We
demonstrate that in strongly confined quantum dots the dark
excitons, which are energetically well below the bright excitons,
have actually a decisive effect on the polarization dynamics due
to their persistent nature. The linear polarization shows
nonexponential decay controlled by a conversion of the dark into a
bright exciton. To get quantitative answers for specific quantum
dot structures, all the necessary information can be obtained
already from experiments on the luminescence dynamics following
nonresonant excitation in these dots.


\end{abstract}

\maketitle

\draft


\section{Introduction}

The dynamics of the spin--polarized excitons in semiconductor
quantum dots (QDs) attracted a great attention from both
fundamental and technological point of view during the two
previous decades. The spin relaxation of excitons between the
bright and dark states strongly limits the generation of single
photons from the dots\cite{Michler,Strauf, Reischle}, whereas the
spin--flip transitions within the bright exciton doublet,
so--called longitudinal spin relaxation, limit the performance of
entangled photon pair generators\cite{Akopian,Stevenson}. So
according to the numerous time--resolved experiments, an inclusion
of dark excitons to the relaxation processes essentially modifies
the luminescence dynamics following nonresonant excitation,
whereas the longitudinal spin relaxation leads to a decay of the
luminescence polarization. Theoretical examinations identified two
relevant microscopic mechanisms of an intrinsic nature responsible
for thermally activated spin--flip transitions within the ground
state of the heavy--hole exciton confined in a QD. The first
mechanism is mediated by the exchange
interaction\cite{Kalt2003,me, Roszak} and the second mechanism is
governed by the spin-orbit coupling\cite{Kalt2005}. For
self--assembled QDs in vertical field effect structures, a
Kondo--like interaction between a localized electron in a dot and
a delocalized electron in the back contact was proposed as the
main microscopic mechanism of the dark--to--bright spin flip in
Ref.\cite{Dal}. Recently, an additional depolarization source due
to transitions between the bright states via the dark states has
been identified in Ref.\cite{Kalt2010}, so that the dark--to
bright transitions also can limit the performance of entangled
photon pair generators. An impact of this relaxation channel on
the polarization dynamics, however, was still not examined in
details. The only exception here, as far as we know, is the
theoretical paper\cite{Roszak}, where the reported results are
based on the numerical calculations for the specific case of the
InAs QDs.

In this paper we present the analytical results on the
polarization dynamics in a single QD solving the phenomenological
rate equations similar to those considered in Ref.\cite{Roszak}.
The paper is organized as follows. Section II contains the problem
statement and the obtained analytical results are discussed in
section III. In section IV the quantitative estimation of
depolarization effects in various quantum dots studied recently in
nonresonantly excited photoluminescence experiments
\cite{Kowalik,Dal,Hvam,Kummell}   and a summary as well are given.

\section{Four--level model.}

Many experimental works devoted to the exciton dynamics in single
quantum dots report on a biexponential decay of the band edge
exciton luminescence as observed e.g. in CdSe/ZnS QDs in
Ref.\cite{Labeau}. The results are interpreted satisfactory in
terms of a three--level model of a zero exciton ground level and
two mixed fine structure states of the heavy--hole exciton, the
bright state $|+1\rangle$ ($|-1\rangle$) and the dark state
$|+2\rangle$ ($|-2\rangle$), see Fig.1(a). This simple model,
however, contains no interplay between two bright states (spin
doublet), which are assumed to be coupled to the dark states in
pairs, and therefore is not able to describe the impact of the
longitudinal spin relaxation. In fact, because of the reduced
symmetry and anisotropic exchange interaction, the two bright
states are separated in energy and form two linearly and
orthogonally polarized eigenstates $|X\rangle$ and $|Y\rangle$.
The spin--flip transitions $|X\rangle \leftrightarrow |Y\rangle$
between these states directly evidence the longitudinal exciton
spin relaxation. A coupling of {\it both} bright states to
\emph{both} dark states opens an additional channel due to the
sequential process like as $|X\rangle \rightarrow |d\rangle
\rightarrow |Y\rangle$ \cite{Roszak, Kalt2010}.

\begin{figure}
\vspace*{0.5cm}
\begin{center}
\includegraphics[scale=.3]{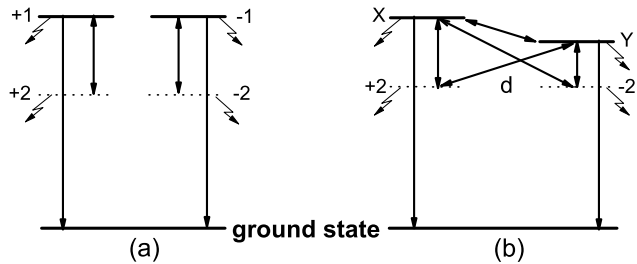}
\caption{Diagram of transitions included in (a) a three--level
model and
 (b) a four--level model.}{\label{Ri1}}
\end{center}
\end{figure}

Based on this scenario we consider a four level model system
containing two bright states $|X\rangle$ and $|Y\rangle$, a
two--fold degenerate dark state \cite{dark}, and the crystal
ground state, as is shown in Fig.1(b). We assume that a short
laser pulse leads to a partial occupation of the involved
excitonic states. Let $N_X(0)$ and $N_Y(0)$ be the population
probabilities (occupations) of the bright states at the time of
initialization $t=0$ and $N_d(0)$ is the joint initial occupation
of the dark state. Supposing that both bright states are coupled
to both dark states \cite{Kalt2010}, further temporal evolution of
the occupations $N_X, N_Y$ and $N_d$ is described by the following
rate equations
\begin{eqnarray} \label{state}
\left.
\begin{array}{lll}
\dot{N_X}(t) &=& - (\widetilde{\gamma_r} + 2 \gamma_{X d} +
\gamma_{XY}) N_X(t) + \gamma_{YX}
 N_{Y} (t) + \gamma_{d X} N_{d}(t), \\
\dot{N_Y}(t)   &=& -(\widetilde{\gamma_r} + 2 \gamma_{Y d} +
\gamma_{YX}) N_{Y}(t)+ \gamma_{XY} N_{X}
(t) + \gamma_{d Y} N_d (t),\\
\dot{N_d}(t) &=& -(\gamma_{nr} + \gamma_{d X} + \gamma_{d Y}) N_d
(t) + 2 \gamma_{X d} N_{X}(t) + 2 \gamma_{Y d} N_{Y}(t),
\end{array}
\right\}
\end{eqnarray}
where $\widetilde{\gamma_r} = \gamma_r + \gamma_{nr}$, $\gamma_r$
is the radiative recombination rate for the bright exciton and
$\gamma_{nr}$ simulates any nonradiative loss (usually the same
for both the bright exciton and the dark exciton). The rates
$\gamma_{i d}$ and $\gamma_{d i}$, $i=X,Y$, characterize the
transitions between the bright and dark exciton states and
$\gamma_{XY}, \gamma_{YX}$ are the transition rates between the
bright states. For the dark exciton, which is optically forbidden
(in the dipole approximation), the radiative recombination rate is
taken to be zero.  As noted above, a similar system of the rate
equations, but with $\gamma_{nr}=0$, was calculated numerically
for a specific case in Ref.\cite{Roszak}.

Below we will restrict our consideration by strongly confined QDs
(like e.g. self--assemble InGaAs/GaAs QDs) for which the reported
values of the bright--dark splitting $\Delta_0$ \cite{Bayer} are
typically (at least) a few times larger than those for the
anisotropic splitting $\Delta_{XY}$ between the bright states
\cite{Kowalik}. One can assume therefore that the spin--flip rates
for both bright excitons must be close in magnitude and further
simplification of the above system is possible.  In what follows
we take $\gamma_{X d}=\gamma_{Y d}\equiv \gamma_{bd}/2$ and,
similarly, $\gamma_{d X} = \gamma_{d Y}\equiv\gamma_{db}/2$.
Excluding also a very low temperature range, where $kT \ll
\Delta_{XY}$, we can take
$\gamma_{XY}=\gamma_{YX}\equiv\gamma_{L}/2$. Finally, it is
convenient to replace the occupations $N_X$ and $N_Y$ by the joint
occupation $N_{+} = N_X + N_Y$ and the occupation difference
$N_{-} = N_X - N_Y$. Under the above approximations the equation
system (\ref{state}) splits into the equation for $N_-(t)$ and two
coupled equations for $N_+(t)$ and $N_d(t)$:
\begin{eqnarray} \label{sum}
\left.
\begin{array}{lll}
\dot{N_-}(t) &=& - (\widetilde{\gamma_r} +
\gamma_{bd}+\gamma_L) N_{-}(t)\;, \\
\dot{N_+}(t)  &=& - (\widetilde{\gamma_r} +
\gamma_{bd}) N_+(t) + \gamma_{db} N_d(t)\;,\\
\dot{N_d}(t)  &=&- (\gamma_{nr} + \gamma_{db}) N_d(t) +
\gamma_{bd} N_+(t)\;.
\end{array}
\right\}
\end{eqnarray}
Solving the above equations, for the occupations of the bright
states we get
\begin{eqnarray}\label{Nx}
N_{X, Y}(t) &=& F_1 \;e^{-\gamma_1 t} + F_2 \;e^{-\gamma_2 t} \pm
F_3 \; e^{- (\widetilde{\gamma_r} + \gamma_{bd}+\gamma_L) t}\;.
\end{eqnarray}
Here the decay rate $\gamma_{1,2} = 0.5 [(\gamma_r + 2 \gamma_{nr}
+\gamma_{bd} + \gamma_{db}) \pm \delta]$ with $\delta =
\sqrt{\Gamma^2 + 4 \gamma_{db} \gamma_{bd}}\;$ and $\Gamma =
\gamma_r + \gamma_{bd} - \gamma_{db}$. The amplitudes $F_{1,2} =
\bigl[(\delta \pm \Gamma)\; N_+(0) \mp 2 \gamma_{db}\;
N_d(0)\bigr]/4\delta$ and $F_3 = N_-(0)/2$. Similarly, the
occupations of the dark states $N_d(t) = D_1 \;e^{-\gamma_1 t} +
D_2 \;e^{-\gamma_2 t}$ with $D_{1,2} = \bigl[(\delta \mp \Gamma)\;
N_d(0) \mp 2 \gamma_{bd}\; N_+(0)\bigr]/4\delta$.

The above results have a simple interpretation. {\bf (i)} Let all
involved excitonic states are initially equally populated - the
situation realized at nonresonant excitation. Then the difference
$N_-(0)=0$ and the initial condition $N_X(0)=N_Y(0)$ persists over
time, that is both radiative excitons decay identically and show
the two--component structure - the result which is typical for a
three level model.
{\bf (ii)} Let now only one bright state be initially populated,
that is $N_-(0)=1$ or $N_-(0)=-1$. This situation is realized at
resonant (linearly polarized) excitation. In this case the
amplitude of the third component for $N_X$ and $N_Y$ differs from
zero,
so that the initially established population is redistributed over
time within the fine--structure levels. This ultimately induces a
depolarization of the resulting emission. The degree of the linear
polarization of the emission is directly related to the
occupations by $P(t) = N_-(t)/N_+(t)$.

\section{Degree of linear polarization}

Let us now consider a short pulsed laser excitation at $t=0$
leading to a population of one of the bright states, say the
$|X\rangle$ state, so that $N_X(0)=1$, but $N_Y(0)=N_d(0)=0$.
According to result (\ref{Nx}), the degree of linear polarization
of the emitted signal is given by
\begin{eqnarray} \label{Pol}
P(t) =  \frac{2 \delta e^{- \frac{1}{2}(\Gamma + 2 \gamma_L)
t}}{(\delta + \Gamma)\; e^{-\delta t/2} + (\delta - \Gamma)\;
e^{\delta t/2}}\;.
\end{eqnarray}
The obtained polarization decay is entirely caused by
fine--structure effects. Sole direct spin--flip transitions $|X
\rangle \leftrightarrow |Y\rangle$ within the radiative doublet
lead to a partial redistribution of the occupation between these
states, but the dark exciton states are, evidently, unaffected.
This is the familiar case of an exponentially decaying
polarization $P(t)= e^{-\gamma_L t}$. The sequential process like
as $|X\rangle \rightarrow |d\rangle \rightarrow |Y\rangle$,
however, causes a nonexponential decay of the joint occupation of
the bright states and effectively contributes to a decay of the
linear polarization. This fact is clearly reflected in the
time--integrated polarization for which we get
\begin{eqnarray}\label{PolAv}
\overline{P} =  \frac{\widetilde{\gamma_r}  + \gamma_{bd} -
 \gamma_{L}^{\star}}{\widetilde{\gamma_r} + \gamma_{bd} +
 \gamma_L}\;.
\end{eqnarray}
Here the involved quantity $\gamma_L^{\star} = \gamma_{bd}
\gamma_{db}(\gamma_{nr} + \gamma_{db})^{-1}$ is interpreted as the
rate of the above sequential transition between the bright states.
Such a transition can be regarded as a second--order (or indirect)
relaxation process, where the dark state serves as the
intermediate state \cite{Kalt2010}. The sequential spin relaxation
is, evidently, the more effective the faster is a conversion of
the dark into a bright exciton in comparison with the nonradiative
relaxation. In a strong conversion regime, where $\gamma_{db} \gg
\gamma_{nr}$, the sequential transition rate reaches the values
$\gamma_L^{\star}\simeq \gamma_{bd}$, whereas $\gamma_L^{\star}
\ll \gamma_{bd}$ in a weak conversion regime, where $\gamma_{bd}
\ll \gamma_{nr}$. Note that a relation between the rates
$\gamma_{db}$ ($\gamma_{bd}$) and $\gamma_{nr}$ can be examined in
experiments on the luminescence dynamics following nonresonant
excitation \cite{Hvam}.

Let us now rewrite the expression (\ref{PolAv}) for the integrated
polarization in a more convenient form $\overline{P} =
\widetilde{\gamma_r} (\widetilde{\gamma_r} + \gamma_s)^{-1}$,
where $\gamma_s = \widetilde{\gamma_r}(\gamma_L +
\gamma_L^{\star}) (\widetilde{\gamma_r} + \gamma_{bd} -
\gamma_L^{\star})^{-1}$ can be viewed as an \emph{effective}
polarization decay rate. In the special case of no bright--dark
coupling the rate $\gamma_s$ is identical to $\gamma_L$, the
polarization decays exponentially, and the definition
$\overline{P} = \widetilde{\gamma_r}/(\widetilde{\gamma_r} +
\gamma_L)$  is the standard one. The more realistic case of an
impact of the dark states complicates the situation. The
(effective) decay rate now has an intricate form and, as could be
expected, is proportional to the total rate of the longitudinal
spin relaxation, $\gamma_s \sim (\gamma_L + \gamma_L^{\star})$.

For the spin relaxation of an intrinsic nature further
simplification of the above results is possible. In this case the
spin--flip transitions are thermally activated and the bright
exciton relaxes predominantly to the lowest dark state \cite{int}.
This means that direct transitions $|X \rangle \leftrightarrow
|Y\rangle$ are slow compared to the bright--to--dark relaxation
and therefore can be abandoned in the analysis, so that we can
chose $\gamma_L=0$ in the first of the equations (\ref{sum}) and,
consequently, in expression (\ref{Pol}). An effective decay rate
in this case is approximated by
\begin{eqnarray}\label{PolApprox}
\gamma_s = \frac{\widetilde{\gamma_r} \gamma_L^{\star}}{
\widetilde{\gamma_r} + \gamma_{bd} - \gamma_L^{\star}} \;.
\end{eqnarray}
It is remarkable that all the dot parameters necessary to estimate
depolarization effects in undisturbed QDs can be deduced in
nonresonantly excited photoluminescence experiments. It follows
immediately from (\ref{PolApprox}) that no noticeable effects have
to occur in a weak conversion regime, where all the involved rates
$\gamma_{bd}$, $\gamma_L^{\star}$ and $\gamma_s$ as well are much
smaller than $\gamma_{nr}$. Respectively, the time--integrated
polarization preserves its initial value, $\overline{P} = 1$. In a
strong conversion regime, where as we saw above
$\gamma_L^{\star}\simeq \gamma_{bd}$, an efficiency of the
polarization decay and the actual values of $\overline{P} =
\widetilde{\gamma_r}(\widetilde{\gamma_r} + \gamma_{bd})^{-1}$ as
well depend on the ratio $\gamma_{bd}/\widetilde{\gamma_r}$.  For
typical self--assembled QDs at low temperatures, intrinsic
spin--flip transitions are much slower than the radiative
recombination, as estimated theoretically \cite{Roszak, Kalt2010}
and observed experimentally, see e.g. \cite{Hvam}. Hence
noticeable effects here can be expected only at enough high
temperatures. Note that self--assembled QDs in a vertical field
effect structures, similar to those studied in Ref.\cite{Dal}, can
also serve as an example of the QDs systems for which the
bright--to--dark processes are preferential. A strong dependence
of the characteristic rate $\gamma_{db}$ on a voltage in these
dots rules out an intrinsic spin flip mechanism, see discussion
below in section IV{\it c}.

\section{Quantitative results}

In the previous section we observed that the experimental
examination of the luminescence dynamics following nonresonant
excitation in QDs presents sufficient information to make a
conclusion about depolarization effects in these dots. It is
possible to get quantitative answers on the polarization decay
rate in such QDs for which the nonresonant experimental data are
at hand and therefore can be used as input. It has to be
implemented for many QD systems in the strong confinement limit
and below we consider several examples.

\paragraph{Self--assembled In(Ga)As QDs.}
A recent series of works \cite{Hvam} reports on the experimental
investigation of nonresonantly excited luminescence dynamics in
self--assembled In(Ga)As QD structures. Actually the following
characteristic rates were deduced from experiments at 14 K:
$\gamma_r \simeq \rm 1\;ns^{-1}$, $\gamma_{nr} \simeq 0.1\;\rm
ns^{-1}$, and $\gamma_{db} \simeq 0.01\;\rm ns^{-1}$. Although no
data on $\gamma_{bd}$ and  $\Delta_0$ as well are reported, we can
consider that $\gamma_{bd} \simeq \gamma_{db}$ ($\Delta_0$ in InAs
QDs is typically of a few hundreds of $\mu$eV \cite{Bayer},
whereas $kT = 1.2\;$meV at 14 K) \cite{rem}. As a result, the
(effective) polarization decay rate is calculated from
(\ref{PolApprox}) to be $\gamma_s \simeq 10^{-3}\;\rm ns^{-1}$, so
that we can conclude that In(Ga)As QDs examined in Ref.\cite{Hvam}
shouldn't show any depolarization effects for the reported
experimental conditions. This conclusion is an accordance to
earlier experimental results from Ref.\cite{Paillard}, where
strictly no decay of the linear luminescence polarization was
observed in self--organized InAs/GaAs QDs at low temperature.
\begin{figure}
\vspace*{0.5cm}
\begin{center}
\includegraphics[scale=.3]{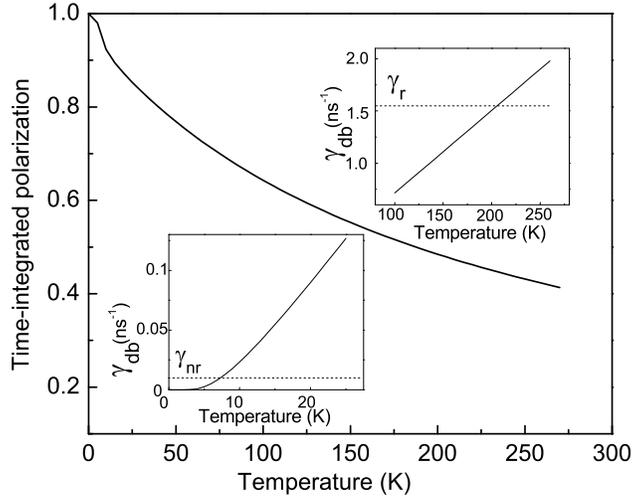}
\caption{Temperature dependence of the time--integrated
polarization and the rate of a conversion of the dark to bright
exciton at low (lower inset) and high (upper inset) temperatures
in QDs from Ref.\cite{Kummell}.}{\label{Ri2}}
\end{center}
\end{figure}

\paragraph{Self--assembled CdSe/ZnSSe/MgS QDs.}
Recently the high--quality CdSe/ZnSSe/MgS self--assembled QDs were
proposed in Ref.\cite{Kummell} as an excellent model system to
analyze experimentally the dark--bright interplay in a wide range
of temperatures. The (nonresonantly excited) luminescence dynamics
investigated in \cite{Kummell} is well explained by a three--level
model and no modification of the nonradiative losses was found all
the way up to room temperature. The following parameters were
deduced from the experiment: $\gamma_r = 1.54 \;\rm ns^{-1}$,
$\Delta_0 \simeq 1.8$ meV, $\gamma_{bd}(\rm 0K) = \gamma_0 = 0.166
\;\rm ns^{-1}$ \cite{rem}, and $\gamma_{slow}({\rm 7K}) \simeq
\gamma_{nr} + \gamma_{db}({\rm 7K}) = 0.02 \;\rm ns^{-1}$ (the
rate of the slow luminescence component at 7 K). Using these data
we extract the nonradiative relaxation rate to be $\gamma_{nr}
\simeq 0.01 \;\rm ns^{-1}$. We are able also to calculate the rate
$\gamma_{db}$ at different temperatures and the result is
presented in the insets in Fig.2. It can be seen that
$\gamma_{db}$ is one order of magnitude larger than $\gamma_{nr}$
at $T \gtrsim 20$ K and becomes comparable (or even larger than)
to $\gamma_r $ at $T \gtrsim 200$ K. We can conclude therefore
that at $\rm T \gtrsim 20$ K the CdSe QDs examined in
Ref.\cite{Kummell} are in a strong conversion regime and must show
a noticeable decay of the linear polarization at high
temperatures. To illustrate the expected depolarization effect, in
Fig.2 we plot the integrated polarization $\overline{P} =
\widetilde{\gamma_r} (\widetilde{\gamma_r} + \gamma_s)^{-1}$ (with
$\gamma_s$ calculated from (\ref{PolApprox})) in a wide
temperature range up to room temperature. It can be seen that the
values of $\overline{P}$ are estimated to be close to unity at
very low temperatures,  but $\overline{P} \simeq 0.87$ is
calculated already at 20 K and $\overline{P} \simeq 0.41$ is found
at 270 K. The experimental proof of these theoretical results is
certainly of interest.

\paragraph{Self--assembled In(Ga)As QDs in vertical field effect structures.}
In recent work \cite{Dal} nonresonantly excited photoluminescence
was measured from self--assembled InGaAs QDs  in a vertical field
effect structure. It was discovered that the rate $\gamma_{db}$ of
a conversion of the dark to bright exciton depends on the applied
voltage and can be tuned to be smaller or larger than the
recombination rate. It was found also that this rate increases
with an increase of the temperature and the emission energy as
well. These experimental findings were successfully explained in
Ref.\cite{Dal} in the framework of a Kondo--like tunneling
interaction between a localized electron in a dot and a
delocalized electron in the back contact.

To draw a conclusion concerning the polarization dynamics in dots
from Ref.\cite{Dal}, note that a strong dependence of the rate
$\gamma_{db}$ on a voltage rules out an intrinsic spin flip
mechanism,  as reported in Ref.\cite{Dal}. It is evident also that
an efficient tunneling mechanism including a simultaneous flip of
the electron spin and the heavy--hole spin is unlikely, so that it
is possible to suppose that still a conversion of the bright to
dark exciton is faster than direct transitions between the bright
states, as in the case of the spin relaxation of an intrinsic
nature. Accordingly, the linear polarization dynamics in dots
examined in Ref.\cite{Dal} can be evaluated from the above
expressions (\ref{Pol}) and (\ref{PolAv}), where we again can
chose $\gamma_L=0$ \cite{tun}. Note also that for a tunneling
mechanism the relation $\gamma_{bd} = \gamma_{db} e^{\Delta_0/kT}$
is still valid \cite{Dal}.

{\bf Table I. In(Ga)As QDs parameters after Ref.\cite{Dal}.}

\begin{tabular}{|c|c|c|c|c|}
 \hline
 & dot{\bf 1} (5 K)  & dot{\bf 2} (5 K) & dot{\bf2} (12 K) & dot{\bf 2} (25 K) \\ \hline
 $E_{\rm PL}$ (eV) & 1.318 &   1.305 & 1.305 & 1.305 \\
 \hline
 $\Delta_0$ (meV) & 0.3 &  0.6 & 0.6 & 0.6 \\ \hline
 $\tau_{r}$ (ns) & 0.55 &   0.77 & 0.77 & 0.77 \\ \hline
 $\tau_{nr}$ (ns) & 16.66 &   16.66 & 16.66 & 16.66 \\ \hline
 $\tau_{db, \;\rm center}$ (ns) & 120 &  110 & 20 & 2 \\ \hline
\end{tabular}

\vspace{0.3cm}

The quantitative results obtained in Ref.\cite{Dal} for two
examined InGaAs dots differing by the exchange splitting
$\Delta_0$, the radiative lifetime $\tau_r = \gamma_r^{-1}$ and
the spin--flip time $\tau_{db} = \gamma_{db}^{-1}$, which we
denote below as dot{\bf1} and dot{\bf2}, are collected in Table I.
The characteristic time $\tau_{nr} = \gamma_{nr}^{-1}$ and the
values $\tau_{db, {\rm center}}$ refer to the bias voltage applied
in the center of the exciton plateau. The temperature at which the
experiment was performed is shown in parentheses.
\begin{figure}
\vspace*{0.5cm}
\begin{center}
\includegraphics[scale=.3]{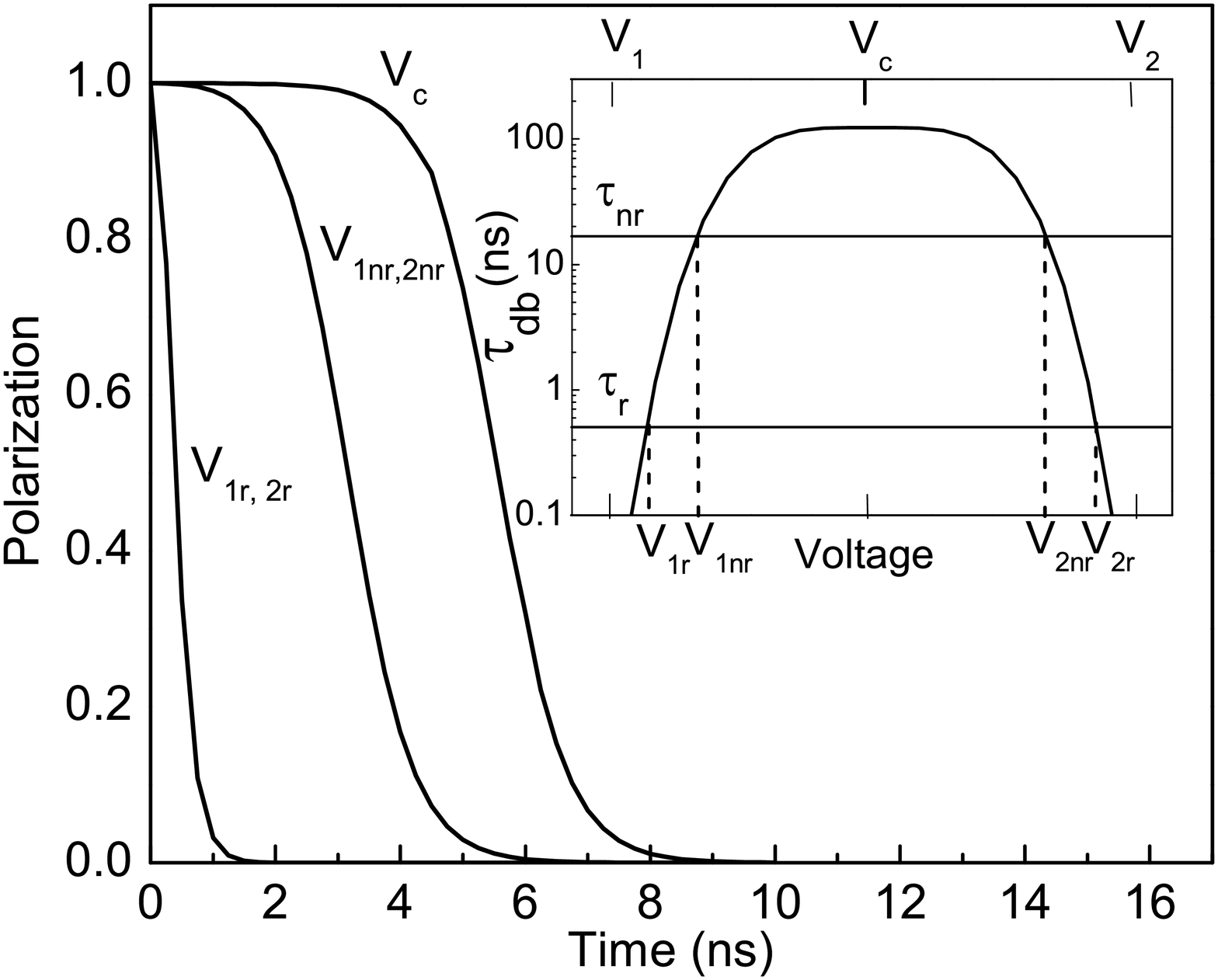}
\caption{The degree of polarization of the emission in the quantum
dot from Ref.\cite{Dal} as a function of time for various voltages
as shown. The inset shows schematic dependence of the
dark--to--bright relaxation time on a voltage within the neutral
exciton plateau with the low--bias and the high--bias edge $V_1$
and $V_2$, respectively, after Ref.\cite{Dal}.}{\label{Ri3}}
\end{center}
\end{figure}
In Fig.3 we plot the evolution of the linear polarization in
dot{\bf1} for various voltages within the exciton plateau as shown
in the inset. The curves are evaluated by means of the expression
(\ref{Pol}) with $\gamma_L=0$. The involved rate $\gamma_{db}$ is
assumed to be $\gamma_{db} = \tau_{db,\; {\rm center}}^{-1}$ for
the curve $V_c$,  $\gamma_{db} = \tau_{nr}^{-1}$ for the curve
$V_{1 nr, 2 nr}$ and finally $\gamma_{db} = \tau_{r}^{-1}$ for the
curve $V_{1 r, 2 r}$, so that all the necessary parameters are
presented in Table I. In Fig.3 is seen that the polarization shows
no decay on a scale of the radiative lifetime $\tau_r$ in the
central part of the exciton plateau. However, the effect becomes
considerable towards either the right or the left edge of the
plateau (the curve $V_{1 r, 2 r}$), where the polarization decay
time $\tau_s=\gamma_s^{-1} = 0.3$ ns (as estimated from
(\ref{PolAv}) with $\gamma_L=0$) is nearly two times smaller than
$\tau_r = 0.55$ ns \cite{lum}. The respective value of the
time--integrated polarization $\overline{P}_{\rm dot{\bf1}}(V_{1r,
2r}) = 0.36$ markedly differs from unity. According to the
experimental data presented in Table I for dot{\bf2}, elevating
temperature strongly accelerates the dark--bright transitions: the
reported values of $\tau_{db,\; {\rm center}}$ are two order of
magnitude larger than $\tau_{r}$ at 5 K, but become comparable to
$\tau_{r}$ at 25 K. Consequently,  at moderate temperatures a
noticeable effect on the polarization dynamics can be expected
already in the center of the plateau. Calculating the
time--integrated polarization from expression (\ref{PolAv}) with
$\gamma_L=0$ and using the data from Table I as input, we get
$\overline{P}_{\rm dot{\bf2}}(V_{c}) = 0.997$ at 5 K, whereas
$\overline{P}_{\rm dot{\bf2}}(V_{c}) = 0.71$ at 25 K.

In the context of excitonic spin relaxation in vertical field
effect structures we would like to address Ref.~\cite{Kowalik}.
Here the time--integrated polarization was investigated in
individual InAs/GaAs self--assembled QDs. Experiments were
performed at $\rm T = 2$ K and the applied voltage was fixed at
the value lying between the center and the high--bias edge of the
neutral exciton plateau. The reported values of the characteristic
time determining the time--integrated polarization for three
examined isolated QDs (with a slightly different emission energy)
range from 10 ns to 20 ns. These values are one order of magnitude
longer than the radiative lifetime reported to be $\tau_r=0.85$ ns
\cite{Kowalik}. Still they are too small to be of an intrinsic
nature, as we saw already at the beginning of this section. As an
alternative version, the above discussed scenario of the
voltage--controlled polarization dynamics can be proposed. This
suggestion, of course, requires experimental verification.

In summary, the characteristic times of depolarization of the
luminescence in strongly confined QDs are mainly controlled by a
decay of the dark exciton and become comparable to the
bright--to--dark relaxation times  at negligible nonradiative
losses. Numerical estimations with the relevant experimental data
on the luminescence dynamics following nonresonant excitation
demonstrate no depolarization effects in self--assembled InAs
quantum dots at low temperatures. Similarly, the quantitative
calculations indicate an efficient polarization decay in the
high--quality self--assembled CdSe QDs at high temperatures and a
voltage--dependent polarization dynamics in self--assembled
In(Ga)As QDs in vertical field effect structures as well.

\section{Acknowledgements} This work was supported by the Center
of Functional Nanostructures (CFN) of the Deutsche
Forschungsgemeinschaft (DFG).


\begin{references}
\bibitem{dark}
The dark states $|\pm 2\rangle$ are hybridized and slightly split
by the anisotropic (short--range) exchange interaction
\cite{Bayer, IvPik}.
\bibitem{rem}
The thermally activated transition rates from the bright to dark
exciton and from the dark to bright exciton are $\gamma_{bd} =
\gamma_0 (N_B + 1)$ and $\gamma_{db} = \gamma_0 N_B$, where $N_B =
1/[\exp{\Delta_0/k_BT}-1]$ is the Bose distribution and $\gamma_0$
is the zero temperature rate of the bright to dark exciton
transition.
\bibitem{tun}
Tunneling--induced coupling of both bright states to the same dark
state seems to be possible.
\end{references}
\end{document}